\newlength\gsfigwidth
\newlength\figwidth
\documentclass[apl,twocolumn,floatfix,amsmath,amssymb,showpacs,showkeys]{revtex4}

\newcommand{\be}{\begin{equation}}
\newcommand{\ee}{\end{equation}}
\newcommand{\ba}{\begin{align}}
\newcommand{\ea}{\end{align}}
\newcommand{\bn}{\begin{eqnarray}}
\newcommand{\en}{\end{eqnarray}}

\usepackage{graphicx}

\begin{document}

%\preprint{Gon\c{c}alves et al.}
\title{Naked-eye visualization of geometric frustration effects in macroscopic spin ices}

\author{R. S. Gon\c{c}alves}
\affiliation{Laboratory of Spintronics and Nanomagnetism ($LabSpiN$),
Departamento de F\'{i}sica, Universidade Federal de Vi\c{c}osa,
36570-900 - Vi\c{c}osa - Minas Gerais - Brazil.}
\author{A. C. C. Gomes}
\affiliation{Laboratory of Spintronics and Nanomagnetism ($LabSpiN$),
Departamento de F\'{i}sica, Universidade Federal de Vi\c{c}osa,
36570-900 - Vi\c{c}osa - Minas Gerais - Brazil.}
\author{R. P. Loreto}
\affiliation{Laboratory of Spintronics and Nanomagnetism ($LabSpiN$),
Departamento de F\'{i}sica, Universidade Federal de Vi\c{c}osa,
36570-900 - Vi\c{c}osa - Minas Gerais - Brazil.}
\author{C.I.L. de Araujo}
\email{dearaujo@ufv.br}
\affiliation{Laboratory of Spintronics and Nanomagnetism ($LabSpiN$),
Departamento de F\'{i}sica, Universidade Federal de Vi\c{c}osa,
36570-900 - Vi\c{c}osa - Minas Gerais - Brazil.}
\author{F. S. Nascimento} \affiliation{Grupo de F\'{i}sica,
Universidade Federal do Rec\^oncavo da Bahia, 45300-000 - Amargosa - Bahia - Brazil.}
\author{W. A. Moura-Melo}
\email{w.mouramelo@ufv.br}
\affiliation{Departamento de F\'{i}sica, Universidade Federal de Vi\c{c}osa,
36570-900 - Vi\c{c}osa - Minas Gerais - Brazil.}
\author{A.R. Pereira}
\email{apereira@ufv.br}
\affiliation{Departamento de F\'{i}sica, Universidade Federal de Vi\c{c}osa,
36570-900 - Vi\c{c}osa - Minas Gerais - Brazil.}
% % % %
% % %
\begin{abstract}
We study planar rectangular-like arrays composed by macroscopic dipoles (magnetic bars with size around a few centimeters) separated by lattice spacings $a$ and $b$ along each direction. Physical behavior of such macroscopic artificial spin ice (MASI) systems are shown to agree much better with theoretical prediction than their micro- or nano-scaled counterparts, making MASI "almost ideal prototypes" for readily naked-eye visualization of geometrical frustration effects.
\end{abstract}
\flushbottom \maketitle
\thispagestyle{empty}
%\section*{Introduction}
Geometrical frustration underlies important physical
phenomena and new states of matter such as spin liquid and spin ice\cite{Moessner-PhysToday-2006}.
However, it is not easy to study frustration in natural systems for the variety of these materials is relatively scarce and their properties cannot be readily controlled. Thus, to bypass such difficulties one has fabricated artificial systems where geometrical frustration can be controlled on demand. For instance, an artificial spin ice (ASI) consist of a regular array of submicrosized elongated ferromagnet rods where geometrical frustration takes place at the vertices. Besides square array \cite{Wang2006,Moller2006,MolJAP2009,MolPRB2010,Morgan2011}, attention has been also given to triangular\cite{MolTriangular}, rectangular \cite{RectangularASI-1,RectangularASI-2,Loreto}, and a number of 'exotic' lattices\cite{Rougemaille,Zhang2013,ReviewASIgeometries-Nisoli}. Despite the great deal of efforts, there remains a number of questions still lacking satisfatory response. For instance, theoretically predicted configurations, namely the ground-state, is rare to be achieved at every vertex throughout the whole sample of nano- or micro-scaled ASI. Besides structural defects incorporated during the as-grown fabrication processes, the actual dynamics of a given nanoisland is far from being of Ising-like\cite{Wysin1,Wysin2}, as assumed by theoretical modeling based upon Monte Carlo and/or dumbbell simulations.  In turn, geometrical frustration transcends scale so that its effects are expected to emerge at macroscopic phenomena, namely in macroscopic frustrated magnetism \cite{Mellado2012}. Here, we conduct a study upon macroscopic artificial spin ice (MASI) systems, disposing the magnetic bars in square and three distinct rectangular arrangements. Our findings clearly demonstrate that MASI's behavior agrees much better with theoretical predictions than their micro- or nano-scaled counterparts, making MASI's very good prototypes to investigate geometrical frustration in magnetic systems.
%
%-------------------------------------------------------------------------------
\begin{figure}
    \centering
   \includegraphics[width=0.48\textwidth]{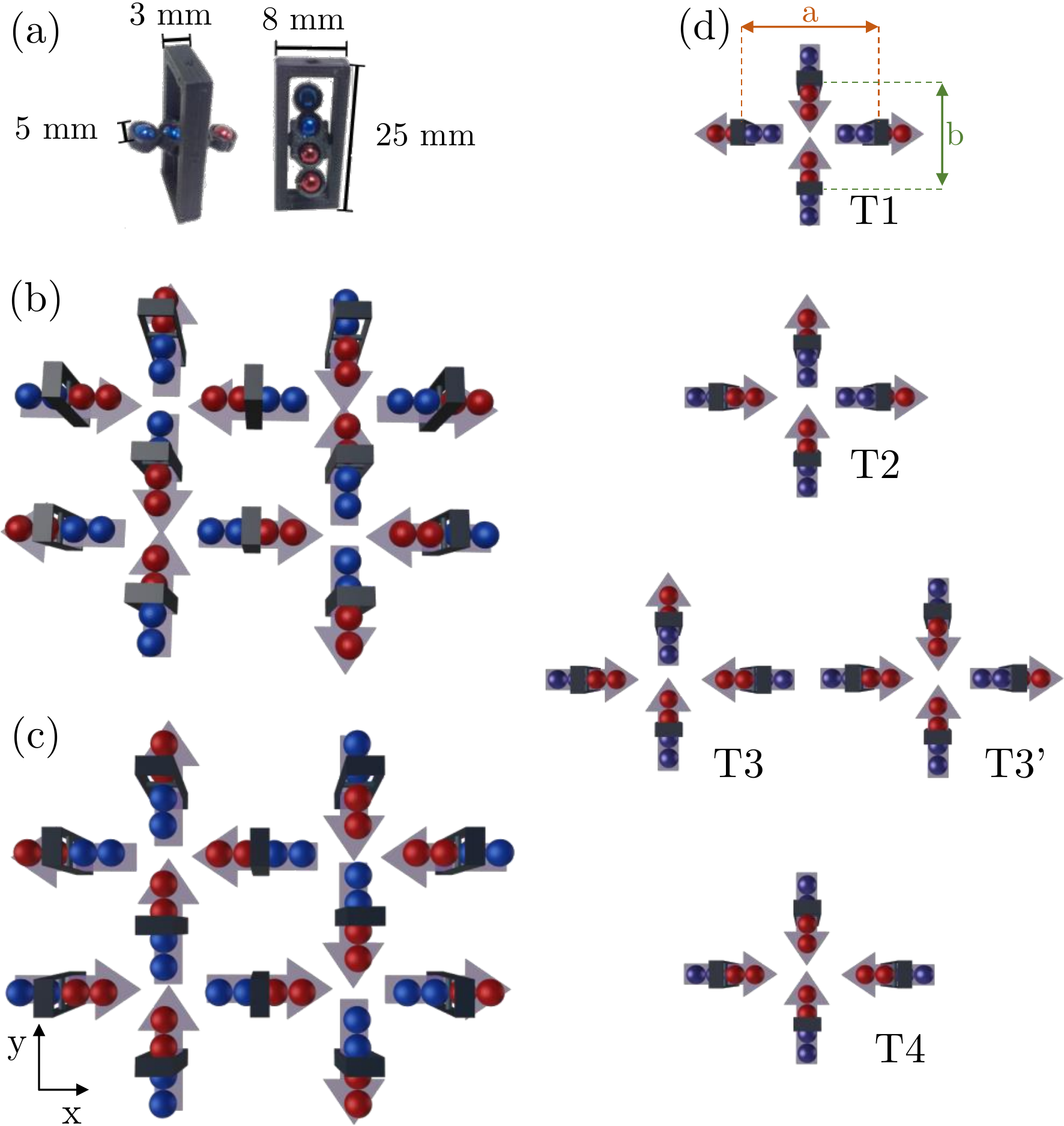}
    \caption{Panel \textbf{(a)} shows how to build MASI's dipoles: Four colored neodymium-made spheres (diameter around 5mm) are disposed to fill hollowed plastic pieces then forming a magnetic bar. Blue and red colors account for the opposite magnetic poles. In panels \textbf{(b)} and \textbf{(c)} the bars are arranged in regular rectangular-like arrays, but keeping dynamics confined to out-of-plane motion. Panel \textbf{(d))} displays the vertex-topologies emerging in such arrays.}
    \label{fig:fig1}
\end{figure}
%-------------------------------------------------------------------------------

Fig. 1 shows how to built MASI samples: in a plastic-made bar (with dimensions 25mm $\times$ 8mm $\times$ 3mm) 4 adjacent holes are filled with colored neodymium-iron-bore alloy, $Nd_2Fe_{14}B$, spheres (with diameter around 5mm and bearing magnetic moment $\sim 10^{-2} {\rm Am^{2}}$) to form a macroscopic magnetic dipole (blue and red spheres represent the opposite magnetic poles). Each bar is constrained to have only out-of-plane dynamics, so behaving as ferromagnetic rotors (friction forces have been diminished at most by construction techniques and suitable lubrication). A set of 42 of such magnetic bars are then disposed in an regular array: 7 rows parallel each other, with 6 bars per row (Fig. 1{\bf c} shows the top view of a typical MASI array). Along $x$-direction, the lattice spacing is fixed at $a=25$mm, while along $y$-axis we have taken 4 distinct spacings: $b=(25; 35.35; 43.3 , 50)$mm, yielding aspect ratios, $\gamma=b/a= (1;\sqrt2;\sqrt3;\sqrt4=2)$.\\

\noindent {\bf MASI arrays and their vertex topologies}: Theoretical calculation for such rectangular arrays predict that the ice regime with degenerate ground state occurs only for $\gamma=\sqrt3$, Refs. \cite{RectangularASI-1,RectangularASI-2}. Indeed, whenever $1
< \gamma < \sqrt{3}$ there are residual magnetic charges at each vertex, alternating from positive to negative at neighbor vertices. In turn, for $\gamma
> \sqrt{3}$ residual magnetic momenta alternate at neighbor vertex. The special case of $\gamma = \sqrt{3}$ comes about for at such a value these two
distinct configurations, vertex with charges and those carrying net magnetization, share the same energy yielding degenerate ground-state. Five possible vertex topologies show up for these lattices, as thoroughly discussed in Ref. \cite{RectangularASI-1,RectangularASI-2}: vertex types $T1$ and $T2$ obey ice rule (two-in, two-out) while the remaining ones represent excited states described as magnetic monopoles: single-charged ($T3$ and $T3 '$) and double-charged poles, $T4$. Fig. 2 also shows the vertex topologies along with ground-state and excited configurations show up in each of the MASI arrays: $R1$ is the square lattice ($\gamma=1$) while $R2$, $R3$ and $R4$ represent the rectangular lattices with
$\gamma=\sqrt{2}$, $\gamma=\sqrt{3}$ and $\gamma=\sqrt{4}$, respectively.\\
% % % % %
\begin{figure}[htp!]
    \centering
    \includegraphics[width=0.48\textwidth]{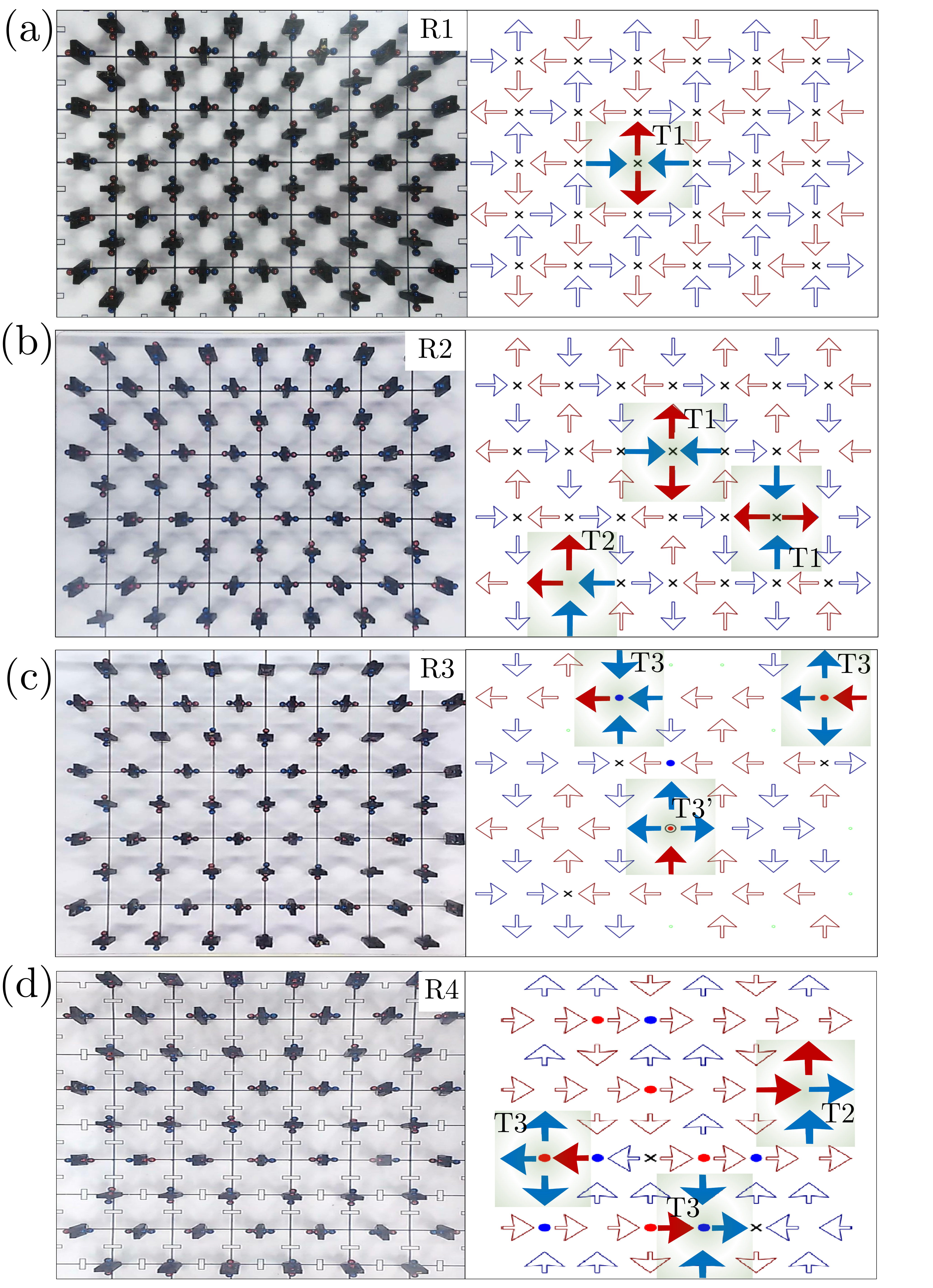}
    \caption {\textbf{(a)}-\textbf{(d)}, left panels show top views of MASI  arrays: $R1$ refers to the square lattice ($\gamma=1$), while the other, $R2$, $R3$, and $R4$ have $\gamma=\sqrt{2}, \sqrt{3},\; {\rm and}\; \sqrt{4}$, respectively. Panels on the right display the respective ground-state of each MASI. The relative population of each vertex type for such MASI ground-states is presented in Fig. 3.}
    \label{fig:fig2}
\end{figure}
% % % % % % % % % % % % % % % %
% % %
% % % % % % % % % % % % % % % % %
\begin{figure}
    \centering
    \includegraphics[width=0.48\textwidth]{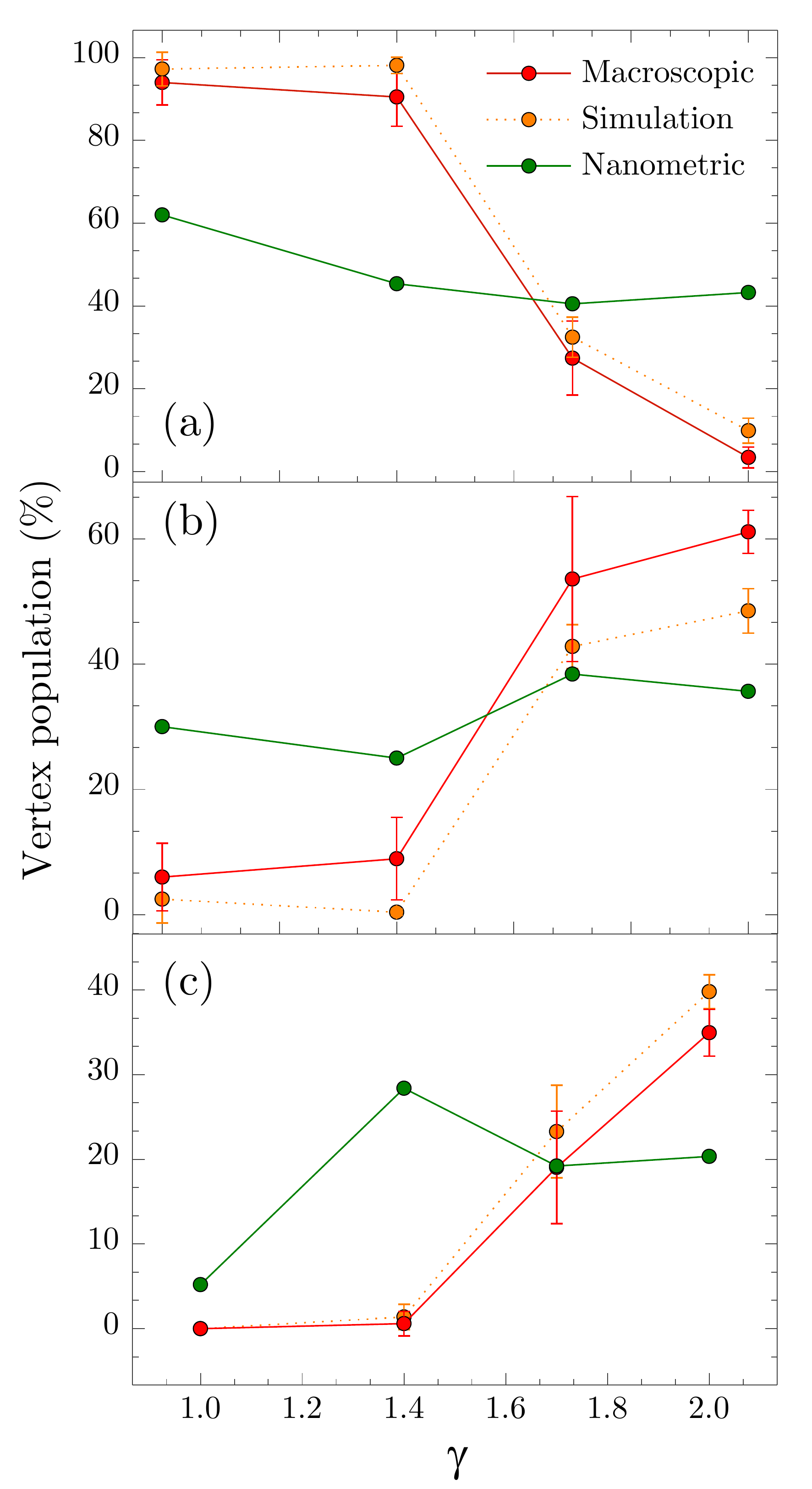}
    \caption{Vertex topologies population vs. aspect ratio, $\gamma$. Each panel shows theoretical/simulation prediction and observed vertex-topology relative population for the ground-state of each macroscopic and nanometric ASI array. Panels {\bf (a)}, {\bf (b)}, and {\bf (c)} correspond to T1, T2, and T3 + T3' populations, respectively. It is noteworthy the much better agreement between prediction and macroscopic spin ice findings than with nanostructured ASI arrays. Experimental results for tiny ASI have been quoted from Refs.\cite{Morgan2011,RectangularASI-2}. [As expected from simulation, $T4$-topology does not show up at all for these ground-states].}
    \label{fig:fig3}
\end{figure}
% % % % % % % % % % % % % % % %

\noindent {\bf Experimental Demagnetization:} Once a given array has been built, a Helmholtz coil pair is used to apply an oscillatory field to the MASI yielding dipoles to experience 'magnetic agitation', which is then realized as mechanical dynamics of the magnetic bars. [Actually, we have realized that even a piece of usual (but strong enough) magnet is capable of doing such a 'magnetic agitation', by zigzag moving it near the array; increasing the frequency even excited states may be achieved]. Thus, a sort of magnetic-induced thermodynamics takes place, sharing similarities to the procedure applied to square ASI as an attempt to achieve ground-state \cite{Wang2006}. After removing this perturbation we have observed that practically all MASI vertex satisfy ice-like rule, $T1$ and $T2$ topologies ($R1$ and $R2$ arrays; the few vertex which have been realized to violate this rule, $\leq 5\%$, are located at MASI borders, where vertex topology breaks down by virtue of dipole-bonds deficit. On the other hand, due to energy degeneracy, $R3$ ground-state bears also a considerable number of $T3$-like vertex. The ground-state of $R4$ has been observed to be highly populated by $T2$ and $T3$ vertices.\\
% %

\noindent {\bf Simulation:} Monte Carlo method has been used to perform the thermodynamics of the rectangular-type spin ice. We consider each dipole as a Ising spin $\vec S_i= \mu s_i \hat{e}_i$, where $\mu$ is the magnetic moment,  $s_i=\pm1$ and $\hat{e}_i$ is the unity vector along  $i$-axis. Each spin is approximated as being a point-like magnetic dipole in such a way that the system is described by the Hamiltonian below:
\begin{equation}
\mathcal{H}= \frac{\mu_0\mu^2}{4\pi} \sum_{i>j} \left[\frac{\hat{e}_i\cdot\hat{e}_j}{r_{ij}^3} - \frac{3(\hat{e}_i\cdot\vec{r}_{ij})(\hat{e}_j\cdot\vec{r}_{ij})}{r_{ij}^5}\right] s_is_j\,, \label{Hamiltonian-model}
\end{equation}
where $r_{ij}$ is the distance between i-th and j-th spin. Open boundary condition is in order so that each spin interacts with all the other remaining ones. In our simulation, arrays with $380$ spins have been used along with $10^4$ Monte Carlo steps to reach equilibrium for every configuration at a 'thermal bath' $T=1.5 D/k_B$ ($D=\mu_0\mu^2/4\pi a^3$ is the magnetic energy scale).

% %
\noindent{\bf Results and Discussion:} Vertex population as function of the aspect ratio, $\gamma$, is presented in Fig. 3. It is noteworthy the much better agreement between theoretical prediction with macroscopic samples than with nano/microsized spin ices. This clearly shows that MASI ground-state is much easier accessed than in tiny systems, where imperfections in nanoislands and its own not-perfectly Ising-like magnetization dynamics are the main facts preventing ground-state achievement throughout the whole sample. [At the borders, where the usual $4$ dipoles per vertex does not hold due to dipole-bonds deficit, it is harder to achieve ground-state, requiring further 'magnetic agitation'].\\
% % % % % % % % % % %
Later, the frequency of the applied field is increased, so that higher power is imputed to the dipoles yielding faster dynamics of the magnetic bars. This enables the system to achieve excited energy states. After some time, around a few dozen seconds, the field is turned off and the remanent excited configurations are in order. Now, monopole pairs connected by strings emerge as the most elementary excitations. An efficient way to realized how much they populate the system is by computing string types, as done in Fig.4. Once more, one notes the better agreement between theoretical prediction and MASI behavior. By virtue of these findings, macroscopic spin ice systems comes to be the best-known ('almost ideal') frameworks to investigate geometrical frustration effects on magnets. Due to its macroscopic scale, they also offer the possibility of naked-eye visualization of such effects, making readily the control of physical properties.
%
%------------------------------------------------------------------------------
\begin{figure}[!h]
    \centering
    \includegraphics[width=0.48\textwidth]{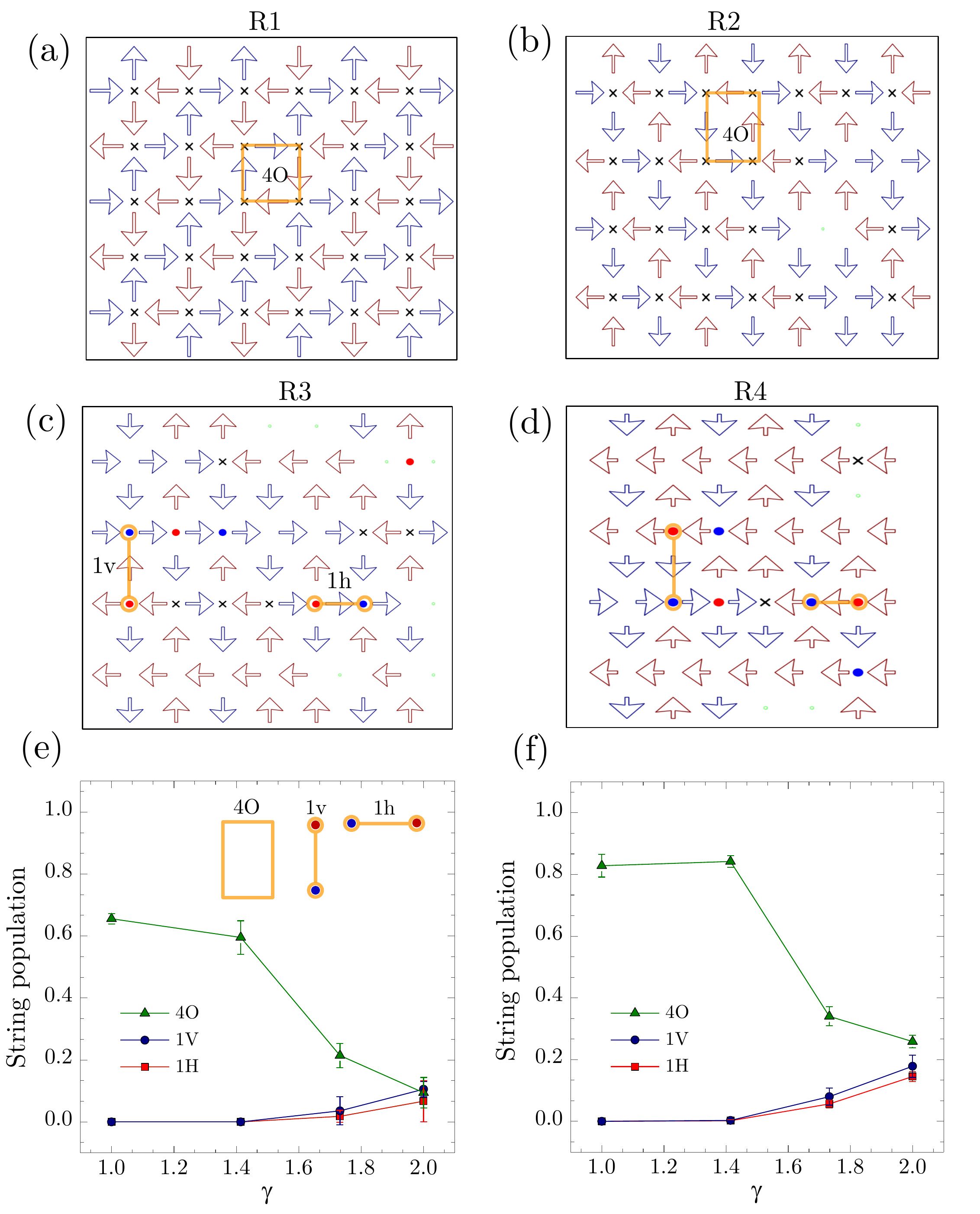}
    \caption{Panels \textbf{(a)}-\textbf{(d)} show schematic representation of the observed strings configurations for the distinct arrays. \textbf{(e)} and \textbf{(f)} depict simulation and MASI observed string-type population with aspect ratio.}
    \label{fig:fig4}
\end{figure}
%------------------------------------------------------------------------------
%
%\section{Concluding Remarks}

\noindent {\bf Conclusions:} Macroscopic artificial spin ice systems are shown to present ground-state and excited configurations populations in very good agreement with theoretical predictions (much better than that verified for micro/nanosized spin ice arrays), suggesting such devices as being the most suitable frameworks for naked-eye visualization of geometrical frustration effects, namely, those encompassed by magnetic interactions.\\

% % %
% % % % % % % % % % % % % %
% % % % % % % % % % % % %
\centerline{\bf Acknowledgements}
The authors thank CNPq, FAPEMIG, and Coordena\c{c}\~ao de Pessoal de N\'{\i}vel Superior- Capes, financial code 001 (Brazilian agencies) for partial financial support. {\em Brazilian Democracy \& Science at risk!}
% % % % % % % % % % % % % % % % % % %
% % % % % % % % % % % % % % % % %

\end{document}